# High-pressure structural phase transitions in $CuWO_4$

J. Ruiz-Fuertes[1,†,*], D. Errandonea[1,2,*], R. Lacomba-Perales[1,*], A. Segura[1,*], J. González[3,4,*], F. Rodríguez[3,*], F.J. Manjón[5,*], S. Ray[5,*], P. Rodríguez-Hernández[6,*], A. Muñoz[6,*], Zh. Zhu[7], and C. Y. Tu[7]

[1]Departamento de Física Aplicada - ICMUV, Universitat de València, Edificio de Investigación, c/Dr. Moliner 50, 46100 Burjassot, Spain

[2]Fundación General Universitat de València, Edificio de Investigación, c/Dr. Moliner 50, 46100 Burjassot, Spain

[3]DCITIMAC, Universidad de Cantabria, Avda. de Los Castros s/n, 39005 Santander, Spain

[4]Centro de Estudios de Semiconductores, Universidad de los Andes, Mérida 5201, Venezuela

[5]Instituto de Diseño para la Fabricación y Producción Automatizada, Universidad Politécnica de Valencia, Camino de Vera s/n, 46022 Valencia, Spain

[6]Departamento de Física Fundamental II, Instituto de Materiales y Nanotecnología, Universidad de La Laguna, La Laguna 38205, Tenerife, Spain

[7]Fujian Institute of Research on the Structure of Matter, Chinese Academy of Sciences, Fujian, Fuzhou, P.R. China

**Abstract:** We study the effects of pressure on the structural, vibrational, and magnetic behavior of cuproscheelite. We performed powder x-ray diffraction and Raman spectroscopy experiments up to 27 GPa as well as *ab initio* total-energy and lattice-dynamics calculations. Experiments provide evidence that a structural phase transition takes place at 10 GPa from the low-pressure triclinic phase ($P\bar{1}$) to a monoclinic wolframite-type structure (*P2/c*). Calculations confirmed this finding and indicate that the phase transformation involves a change in the magnetic order. In addition, the equation of state for the triclinic phase is determined: $V_0$ = 132.8(2) Å$^3$, $B_0$ = 139 (6) GPa and $B_0'$ = 4. Furthermore, experiments under different stress conditions show that non-hydrostatic stresses induce a second phase transition at 17 GPa and reduce the compressibility of $CuWO_4$, $B_0$ = 171(6) GPa. The pressure dependence of all Raman modes of the triclinic and high-pressure phases is also reported and discussed.

[†] Corresponding author: javier.ruiz-fuertes@uv.es
[*] Member of the MALTA Consolider Team

PACS numbers: 62.50.-p, 61.50.Ks, 61.05.cp, 64.30.Jk, 63.20.dd

## I. Introduction

Since the discovery of high-temperature superconductivity and its relation with the coordination of Cu with O atoms and the linking of these polyhedra to networks, the study of the crystal structure of Cu compounds has experienced an increased interest [1]. Although quaternary oxides are in the centre of interest, the effects of pressure in the Cu environment in systems like $CuWO_4$ could be helpful in the search of routes for superconductivity. Several high-pressure structural studies have been performed in compounds related with $CuWO_4$ [2 - 4]. However, due to its low crystal symmetry, the crystallographic study under pressure has shown to be more complicated.

Copper tungstate ($CuWO_4$, the mineral cuproscheelite) is a member of the wolframite series of structurally related materials. It crystallizes in a triclinic structure ($P\bar{1}$) [5] with both cations octahedrally coordinated by O atoms. The $CuO_6$ octahedra present a Jahn-Teller distortion that gives rise to an approximately elongated octahedron. This causes a distortion of the lattice that produces the 2-fold axes and mirror planes disappearance, lowering the crystal symmetry from *P2/c* (wolframite) to $P\bar{1}$ . Since the refinement of the $CuWO_4$ structure [5], several structural and vibrational studies have been carried out in $(Zn_xCu_{1-x})WO_4$ solid solutions [6 – 9]. It is known that a ferroelastic phase transition from cuproscheelite to wolframite takes place at $x = 0.75$. More recently, optical and Raman measurements [10, 11] provided evidence on the existence in $CuWO_4$ of a phase transition near 10 GPa. However, there is no structural information on the high-pressure phase and only part of the Raman modes have been explored under compression up to 16 GPa. The aim of this work is to study the compressibility and structural phase transitions, as well as further explore the vibrational and magnetic properties of cuproscheelite. In order to achieve this goal we performed angle-dispersive x-ray diffraction (ADXRD) and Raman spectroscopy studies under different stress conditions. The experimental studies are combined with *ab initio* total-energy and lattice dynamics calculations.

## II. Experimental details

We performed two independent x-ray diffraction experiments. In one experiment (exp.1), silicone oil (SO) was used as pressure-transmitting medium. This experiment was performed at beamline I15 in the Diamond Light Source with a monochromatic x-ray beam ($\lambda$ = 0.61506 Å), which was focused down to 30 μm x 30 μm using K-B mirrors. A membrane-type diamond anvil cell (DAC) with 400-μm diameter culet diamonds and a 180-μm drilled Inconel gasket was used. In the other experiment (exp. 2), argon (Ar) was the pressure medium. This experiment was carried out with a symmetric DAC with 480-μm diameter culet diamonds and a 150-μm drilled rhenium gasket, at 16-IDB beamline of the HPCAT at the Advanced Photon Source (APS). In this case, a monochromatic x-ray beam ($\lambda$ = 0.36783 Å) was focused down to 10 μm x 10 μm. In both experiments the diffraction patterns were recorded on a MAR345 image plate located at 450 and 350 mm from the sample, respectively, and integrated using

FIT2D. To perform the experiments we used micron-size powders with purity higher than 99.5% (Mateck). The pressure was measured by means of the ruby fluorescence technique. In the second experiment, pressure was confirmed with the equation of state (EOS) of Ar [12]. Prior to loading the DACs, x-ray diffraction and Raman measurements confirmed that only the low-pressure triclinic phase is present in the samples. The unit-cell parameters were $a$ = 4.709(7) Å, $b$ = 5.845(9) Å, $c$ = 4.884(7) Å, $\alpha$ = 88.3(2)º, $\beta$ = 92.5(2)º, and $\gamma$ = 97.2(2)º, in agreement with Ref. 5. The indexation of the Bragg reflections was done with UNITCELL and DICVOL. GSAS was used to carry out LeBail refinements [13] of the low- and high-pressure structures.

Two Raman studies were performed in 10 μm thick platelets cleaved from a single crystal. The single crystal preparation was described in Ref. 10. For these measurements we used a membrane-type DAC with 500-μm diameter culet diamonds and a 200-μm drilled Inconel gasket. A 16:3:1 methanol-ethanol-water mixture (MEW) and SO were used as pressure-transmitting media. Pressure was determined using the ruby fluorescence scale. In one experiment the Raman spectra was measured up to 17 GPa with the 647.1 nm line of a Coherent krypton laser (model Innova 300) using a Ramanor U1000 double monochromator equipped with a LN refrigerated Symphony CCD detector. The Raman spectra under pressure were obtained in second order with an attached confocal microscope. In the second experiment, measurements up to 21.1 GPa were performed in a backscattering geometry using a LabRam HR UV microRaman spectrometer with a 1200 grooves/mm grating and 100 μm slit, in combination with a thermoelectric-cooled multi-channel CCD detector. A He-Ne 632.81 nm laser line with a power below 10 mW was used for Raman excitation to avoid thermal effects, since the usage of higher laser power showed the appearance of burned areas in the sample. The silicon Raman mode (520 $cm^{-1}$) was used as a reference for the calibration of the Raman spectra. In both measurements the spectral resolution was below 2 $cm^{-1}$.

## III. Calculation technique

Ab initio total-energy and lattice-dynamics calculations were done within the framework of the density-functional theory (DFT) and the pseudopotential method using the Vienna *ab initio* simulation package (VASP) [15], a first principles plane-wave code which can describe the exchange and correlation energy in the local-density approximation (LDA) and in the generalized-gradient approximation (GGA). A detailed account can be found in Refs.16-19. All our calculations are performed at T = 0 K. The exchange and correlation energy was initially taken in the GGA according to Perdew-Burke-Ernzerhof (PBE) prescription [20]. The projector-augmented wave (PAW) scheme [21] was adopted and the semicore 5*p* electrons of W were dealt with explicitly in the calculations. The set of plane waves used extended up to a kinetic energy cutoff of 520 eV. This large cutoff was required to deal with the O atoms within the PAW scheme to ensure highly converged results. The Monkhorst-Pack (MP) [22] grid used for Brillouin-zone integrations ensured highly converged results for the analyzed structures (to about 1 meV per formula unit). We performed spin density calculations and we found that the antiferromagnetic configuration was the stable one for the low-pressure triclinic phase ($P\bar{1}$), whereas the ferromagnetic configuration was the most stable for the monoclinic wolframite-type structure (*P2/c*). At each selected volume, we optimized the atomic geometry including ionic coordinates, the structures were fully relaxed to their equilibrium configuration through the calculation of the Hellman-Feynman forces on atoms and the stress tensor, see Ref. 23. In the relaxed configurations, the forces are less than 0.006 eV/Å and the deviation of the stress tensor from a diagonal hydrostatic form is less than 0.1 GPa. The highly converged results on forces are required for the calculation of the dynamical matrix using the direct force constant approach (or supercell method) [24] .The construction of the dynamical matrix at the Γ point is quite simple and involves separate calculations of the forces in which a fixed displacement from the equilibrium configuration of the atoms within the *primitive* unit cell is considered. Symmetry aids by reducing the number of such independent distortions and reducing the amount of computational effort in the study of the analyzed structures considered in our work. Diagonalization of the dynamical matrix provides both the frequencies of the normal modes and their polarization vectors, it allows us to identify the irreducible representation and the character of the phonon modes at the zone center. In the manuscript we will comment on the pressure dependence of the Raman active modes of the different structures. For completion, we report the frequencies and pressure coefficients of the IR active modes for the triclinic and monoclinic phases as an appendix.

## IV. Results and discussion

### A. X-ray diffraction

In Figs. 1 and 2 we show selected x-ray diffraction patterns of ADXRD experiments 1 and 2 performed with SO and Ar, respectively. In both figures it can be observed that up to 7.8 GPa all the Bragg reflections can be indexed according to the $CuWO_4$ triclinic structure (phase I). Beyond 9 GPa extra peaks appear (depicted by arrows in the two figures) pointing out the onset of a phase transition, to a phase with monoclinic symmetry that we will denote as phase II, in good agreement with previous Raman and optical absorption experiments [11]. As we will explain later, the triclinic structure coexists with the monoclinic phase II during more than 6 GPa after the transition onset. Note that in experiment 2, one peak of the *fcc* structure of Ar is identified at about 2θ = 7.6º [12]. This peak is depicted with a star and is easy to identify since its pressure shift is different from that of the $CuWO_4$ peaks. In experiment 2, we do not find any evidence of additional structural changes or chemical decomposition of $CuWO_4$ up to 20.3 GPa. In contrast, in experiment 1, beyond 16 GPa there is an extinction of the peaks of the triclinic phase and additional peaks are detected. In particular, a new peak located around $2\theta = 7.5^o$ and another one around $2\theta = 11.5^o$ can be clearly seen in the figure at 20.3 GPa. Both facts indicate the onset of a second transition to a phase that we will name phase III. The detected structural changes are reversible in both experiments as can be seen in Figs. 1 and 2. However even though the triclinic phase is fully recovered after pressure release, phase III is present on decompression up to 8 GPa when SO is the pressure-transmitting medium. We will see later that Raman measurements fully support this observation and allow us to confirm that only one phase transition takes place upon decompression from phase III to I.

From the analysis of all the diffraction patterns, we have obtained the pressure behavior of the lattice parameters of the low-pressure phase of $CuWO_4$ (phase I). A LeBail refinement has been used to fit the x-ray diffraction profiles (see Fig. 3 and Table I) and the normalized unit-cell parameters are reported as a function of pressure up to 10 GPa in Fig. 4. We observe that the *c*-axis is less compressible than the other two axes. This anisotropic compression is higher in the experiment performed under Ar. Similar anisotropic behaviors were found also in the structurally related $CdWO_4$, $MgWO_4$, $MnWO_4$, and $ZnWO_4$ [4, 25]. This fact is related to the different linking of octahedral units along different crystallographic directions, being the $CuO_6$ octahedra much more compressible than the $WO_6$ octahedra. Figure 5 shows the pressure dependence of the unit-cell volume which was analyzed using a second-order Birch–Murnaghan EOS ($B_0$' = 4) [26]. The unit-cell volume ($V_0$) and the bulk modulus ($B_0$) at zero pressure obtained for the triclinic phase are: $V_0$ = 132.8(2) Å$^3$ for both experiments, and $B_0$ = 171(2) and 134(6) GPa for experiments 1 and 2, respectively. The fitted EOSs are shown as lines in Figure 5. The obtained bulk moduli agree with the value calculated using the empirical model proposed in Ref. 27, which gives an estimated value of 158 GPa. The bulk modulus of $CuWO_4$ between 139 and 171 GPa is also comparable with those obtained in wolframite-structure tungstates [4].

Our experiments show that $CuWO_4$ is 18% more compressible using Ar than using SO. This medium-dependent behavior found in $CuWO_4$ can be explained if we have into account that Ar is a better hydrostatic medium than SO [14, 28]. It is well known that non-hydrostatic effects can influence the structural properties of a material if its mechanical strength is smaller than the one of ruby (like $CuWO_4$) [29]. Indeed, there are many examples in the literature of ternary oxides (e.g. $ZrSiO_4$, $CaWO_4$) [27, 30] where bulk modulus differences of up to 20% are found depending on hydrostaticity despite being highly uncompressible materials. Usually, larger bulk moduli are systematically obtained with the less hydrostatic media [31] a fact that agrees with our findings for $CuWO_4$. The non-hydrostatic effects could be also responsible for the second phase transition detected in the experiment done under SO, a fact that we confirmed by Raman spectroscopy. It has been observed in tungstates like $BaWO_4$ and $PbWO_4$ that the structural sequence is very sensitive to any degree of non-hydrostaticity [32]. Therefore, apparently non-hydrostatic conditions induced by the use of MEW and SO as pressure-transmitting media [14], make $CuWO_4$ less compressible and favor the occurrence of a second transition. The influence of non-hydrostatic stresses can be also seen in the inset of Fig. 5. There, it is shown that the Bragg peaks considerably broaden under compression when SO is the pressure medium. In addition, changes in the evolution with pressure of the full-width half-maximum (FWHM) of the (010) reflection takes place around 7.5 GPa, supporting that the phase-transition onset occurs near this pressure.

Additional information on the structural high-pressure behavior can be extracted from the study of the three triclinic angles of $CuWO_4$. As shown in Fig. 4, a clear symmetrization is suffered by $CuWO_4$ up to the onset of the phase transition with all the angles getting closer to 90º. Particularly interesting is the behavior of $\alpha$ and $\beta$ angles. Both decrease upon compression taking the same value when the phase-transition onset is detected. All these facts could be related with a symmetrization of the $CuO_6$ octahedra associated to a reduction of the Jahn-Teller (JT) distortion. This phenomenon can be quantified through the pressure effects on the JT distortion parameter, defined as:

$\sigma_{JT} = \sqrt{\frac{1}{6}\sum_{i=1}^{6}\left(R_{Cu-O} - \langle R_{Cu-O}\rangle\right)^2}$, where $R_{Cu-O}$ are the six Cu-O distances of the $CuO_6$ octahedra and $\langle R_{Cu-O}\rangle$ is the average Cu-O distance. From our results, it can be deduced that $\sigma_{JT}$ decreases from 0.201 Å at ambient pressure to 0.160 Å at 10 GPa, approaching the value of $\sigma_{JT}$ in monoclinic wolframite-type $CdWO_4$ (0.095 Å) and $MnWO_4$ (0.088 Å). This symmetrization of the $CuO_6$ octahedra together with the fact that the triclinic structure of $CuWO_4$ is a symmetry-reduced version of wolframite, suggest that the high-pressure phase (phase II) might have a monoclinic wolframite-type structure. This fact is consistent with present and previous experiments [11] which found that the Raman spectra measured in phase II resemble very much those measured for wolframite $ZnWO_4$ and $CdWO_4$ [33, 34].

This hypothesis was used to analyze the diffraction data of phase II. We found that the diffraction patterns of phase II cannot be properly indexed considering only a wolframite-type phase (*P2/c*, Z = 2). However, we have been able to index the diffraction patterns assuming the coexistence of phases I and II (see Fig. 3). The unit-cell parameters obtained for both structures at 16 GPa are given in Table I. Our results indicate that a volume change of about 1% occurs at the triclinic-monoclinic transformation. In view of this evidence, the proposed monoclinic structure appears as the most probable for the high-pressure (HP) coexisting phase. This conclusion is also supported by *ab initio* calculations as we will describe in Section IV.B. The coexistence of both structures is compatible with the domain formation we detected by visual observation in single-crystals. Macroscopic fringes are systematically observed at pressures close to the phase transition as can be seen in Fig. 6. We would like to add here that the proposed structural sequence agrees with the systematic proposed for orthotungstates based upon crystallochemical arguments [27]. The phase transition is also consistent with the fact that in solid solutions of $CuWO_4$ and $ZnWO_4$ ($NiWO_4$) an increase of the Zn (Ni) concentration induces a volume reduction and the transition from cuproscheelite to wolframite at around $Zn_{0.78}Cu_{0.22}WO_4$ ($Ni_{0.6}Cu_{0.4}WO_4$) [6, 35].

Finally, we have to note that no information about the structural symmetry of phase III, observed in the experiment performed with SO, could be obtained from our ADXRD experiments. The broadening of diffraction peaks beyond 16 GPa together with the coexistence of phases precludes any reliable structural identification of phase III.

### B. Raman spectroscopy

In order to complement the x-ray diffraction study of $CuWO_4$ we explored its vibrational properties by means of Raman spectroscopy. Should the proposed phase coexistence be correct, we should observe up to 36 Raman modes, 18 $A_g$ corresponding to the triclinic phase plus eighteen (8 $A_g$ + 10 $B_g$) of wolframite. Figure 7 shows selected Raman spectra of $CuWO_4$ up to 21 GPa. The Raman spectra up to 10.5 GPa correspond to the low-pressure triclinic phase with its 18 Raman active modes [11].

Above this pressure, extra peaks appear whereas those of the low-pressure phase are still present. At 12.5 GPa the number of peaks is already 36 as it was expected. This evidences the onset of a phase transition. Out of the 36 peaks, 18 can be assigned to the triclinic phase and the emerging 18 peaks are consistent with the HP monoclinic wolframite phase detected in x-ray diffraction experiments. These modes that appear after the phase transition resemble very much the Raman spectra of wolframite $CdWO_4$ and $ZnWO_4$ [33, 34]. Lattice-dynamics calculations confirm the assignment of the additional 18 modes detected for phase II to a monoclinic wolframite-type structure. All these facts together with the x-ray diffraction analysis support the identification of the post-triclinic phase as a wolframite structure. The experimental frequencies of the Raman active modes that correspond to phases I and II are indicated with ticks in Fig. 7 at different pressures. Note that additional changes leading to the appearance of sixteen different Raman modes occur in the Raman spectra at 17.1 GPa confirming the occurrence of a second phase transition. Fig. 8 shows a plot of the observed Raman peaks positions as a function of the pressure. A summary of the frequencies (ω) and pressure coefficients of the different modes for the three phases is given in Tables II and III. Table II also reports the Grüneisen parameters ($\gamma = B_0/\omega \cdot d\omega/dP$) for the triclinic phase calculated using the bulk modulus obtained from ADXRD measurements using SO were used ($B_0$ = 171(6) GPa).

We would like to comment here that the coexistence of two phases which has been observed by means of both spectroscopic and structural techniques, could very well explain the previous hypothesis presented in Ref. 11 where a coexistence of Cu atoms in two atomic positions is argued. In that work it is established that both positions would keep the Cu atoms octahedrally coordinated, but while the large Jahn-Teller distortion would remain for one of the positions ($P\bar{1}$ structure) for the other one the Cu atom could be occupying a nearly $O_h$ position (wolframite *P2/c*) giving rise to a very well defined $e_g \rightarrow t_{2g}$ optical absorption at the phase transition.

Regarding the second phase transition occurring above 17 GPa, we have identified up to 16 modes in phase III (see Table III and Fig. 7). On pressure release phase III persists down to 8 GPa where the Raman spectrum of the triclinic phase is recovered. The large hysteresis indicates that the second transition is strongly first order and points out the irreversibility of the phase III → phase II transition. Although we were unable to identify phase III from x-ray diffraction experiments, our Raman results could provide some important information regarding the building block and the coordination of the W-O and Cu-O polyhedra. First, the Raman spectra of phase III do not show any resemblance to those of the post-wolframite phases in $CdWO_4$ and $ZnWO_4$ [33, 34]. In particular, the most intense highest-frequency mode, which corresponds to the totally symmetric W-O stretching vibration, does not drop in frequency as typically occurs at the HP phase transition in wolframites. This strongly suggests that no significant change is expected in the W-O coordination in the phase II – phase III transition, what is a worth information to identify the crystal structure of phase III.

### C. Calculations

To help in the interpretation of our experimental results, *ab initio* total-energy and lattice-dynamics calculations were performed for $CuWO_4$. Along with the triclinic structure ($P\bar{1}$) a monoclinic wolframite-type structure (*P2/c*) was considered. According to the calculations the triclinic phase is the most stable structure from ambient pressure up to 9.3 GPa. Beyond this pressure a wolframite-type structure becomes energetically more favourable which is in good agreement with the phase transition detected in the experiments. For the triclinic structure at ambient pressure, the calculations gave a = 4.8352 Å, b = 6.0538 Å, c = 4.9429 Å, $\alpha$ = 93.26º, $\beta$ = 94.25º and $\gamma$ = 80.11º. The obtained atomic positions are summarized in Table IV. The calculated unit-cell parameters are slightly larger than the experimental values. This overestimation is within the typical reported systematic errors in DFT-GGA calculations. The calculated EOS of phase I is given by the following parameters $B_0$ = 77 GPa, $B_0$' = 4, and $V_0$ = 142 Å$^3$. The obtained bulk modulus is 40% smaller than our experimental value. A possible reason for it is the known fact that GGA typically tends to underestimate the bulk modulus due to approximation used to calculate the exchange-correlation energy [36]. In contrast, LDA calculation gives a bulk modulus of 134 GPa closer to the experimental value; however its description of $CuWO_4$ is very poor. Therefore regarding high-pressure structural stability and phonons we will report the results obtained using GGA. The calculations also indicate that triclinic $CuWO_4$ is magnetically ordered at 0 K having the equilibrium structure an antiferromagnetic (AF) state. Cu cations have a magnetic moment of 0.52 $\mu_B$ in good agreement with previous theoretical calculations [37] and the 0.67 $\mu_B$ experimental value [38].

In addition to the structural calculations we have also performed lattice-dynamical calculations for phase I. Table II compares the calculated frequencies and pressure coefficients for the Raman modes with the experimental ones. The agreement between calculations and experiments for the low-pressure phase is reasonably good and similar to that obtained in $ZnWO_4$ and $CdWO_4$ [33, 44]. This fact supports the lattice dynamics calculations performed for the high-pressure phase of $CuWO_4$. As it normally happens in these compounds the difference between experimental and calculated modes frequencies increases for the higher-energy modes [33, 34].

Considering now phase II, in our experiments we found evidence of a phase transition at 10 GPa and the calculations found that at the same pressure, phase I becomes unstable against a wolframite-type phase (see Fig. 9). The structural parameters of the monoclinic phase are summarized in Table IV at 10.3 GPa. In fact if we compare the calculated lattice parameters a = 4.494 Å, b = 5.581 Å, c = 4.834 Å and β = 89.62º at 16.8 GPa with the experimental values showed in Table I at 16 GPa we observe that the values are very similar. Moreover, according to the calculation in the phase II structure the $CuO_6$ octahedral distortion is very reduced showing a Jahn-Teller distortion of only $\sigma_{JT}^{teo}$(10.3 GPa) = 0.033 Å when $\sigma_{JT}$ = 0 means $O_h$ symmetry. If we compare with the distortion showed by other wolframites and especially with the same arrangement in the triclinic phase we could very well make the approximation of a nearly octahedral coordinated $CuO_6$ with a completely quenched Jahn-Teller distortion in very good agreement with previous work hypothesis [11], as commented earlier in section B. In contrast with the low-pressure phase, according with the calculations, phase II has a ferromagnetic (FM) order at 0 K. In this case Cu cations have a magnetic moment of 0.959 $\mu_B$. This fact implies that the triclinic-monoclinic transition goes beyond a mere structural transition involving also an AF-FM transition at low temperatures. Since strong spin-lattice interactions are determinant in the magnetic order of $CuWO_4$ [39] these interactions should be affected at the transition to allow the observed change of the magnetic order. Note that this change is analogous to findings in $CuMoO_4$ whose monoclinic ε-$CuMoO_4$ phase is FM ordered [40].

From our calculations we also determined the EOS for the HP phase, using a Birch-Murghnagam EOS we obtained the following parameters: $B_0$ = 118 GPa, $B_0^{'}$ = 6.7 and $V_0$ = 135.1 Å$^3$. The calculated bulk modulus, for the wolframite structure of $CuWO_4$, is similar to that of other wolframites and clearly suggests that the wolframite phase is much less compressible than the triclinic phase. Unfortunately we cannot compare this result with the experiments (the phase coexistence observed beyond 10

GPa does not allow to accurately determine the EOS for wolframite). Theoretical calculations suggest a volume reduction of 1.28% from phase I to phase II at the transition pressure.

In relation to the calculated Raman mode frequencies and pressure coefficients for phase II, there are many mode frequencies that match very well the experimental values while for other modes the differences are bigger, especially for higher-frequency modes. The same thing happens for the pressure coefficients but in general the pressure behavior is well predicted. Finally, for phase III no calculations have been carried out since this third phase is only observed under non hydrostatic conditions and it is therefore difficult to be predicted by *ab initio* calculations.

## V. Concluding remarks

In summary, we have carried out two ADXRD experiments in the $CuWO_4$ using two different pressure-transmitting media (SO and Ar) up to 27 GPa and Raman spectroscopy measurements up to 21 GPa using MEW and SO as pressure-transmitting medium. Experimental measurements were complemented with *ab initio* total-energy and lattice dynamics calculations. We have obtained the compressibility of the material in both SO and Ar media showing a higher compressibility and anisotropy under quasy-hydrostatic conditions. In particular, an EOS is reported for $CuWO_4$, being $B_0$ = 139(6) GPa, a value similar to that of wolframite-structure tungstates. The detection of two phase transitions has been also reported at 10 and 16 GPa, with the second one only being detected under non-hydrostatic conditions. A possible structure for the high-pressure phase (phase II) is proposed and confirmed by the calculations, having this phase a monoclinic wolframite-type structure. Additionally, we found that pressure induces a reduction of the Jahn-Teller distortion in $CuWO_4$, an enhancement of the symmetry of the low-pressure phase, as well as a possible Jahn-Teller quenching in the wolframite phase. Raman measurements, confirm the phase transitions observed by x-ray diffraction. We also determine the frequency and pressure dependence of all first-order modes of the low-pressure triclinic and high-pressure phases. This phase transition has shown to be energetically favorable according to calculations, which additionally unravels the occurrence of an AF-FM phase transition together with the structural transformation. On top of that, lattice-dynamics calculations provided information about the Raman and infrared active modes as well as their HP behavior. The structure of the second HP phase (phase III) has not been identified so far, but at least 16 modes have been indentified for it. The Raman spectra of phase III suggest that W coordinated to six oxygen atoms as in phases I and II.

## VI. Appendix

For completion in Table V present the IR active modes for the triclinic and monoclinic phases as well as their pressure coefficients.

**Acknowledgments:** Research financed by the Spanish MEC under Grants MAT2007-65990-C03-01, MAT2007-65990-C03-03 and CSD-2007-00045. Part of the experiments was conducted with the support of the Diamond Light Source at the I15 beamline (proposal No. 683). Portions of this work were performed at HPCAT (16-IDB), Advanced Photon Source (APS), Argonne National Laboratory. HPCAT is supported by DOE-BES, DOE-NNSA, NSF, and the W.M. Keck Foundation. APS is supported by DOE-BES, under Contract No. DE-AC02-06CH11357. The authors thank A. Kleppe and Y. Meng for technical support during the experiments. J. R.-F. and R. L.-P. thank the support from the MEC through the FPI and FPU programs. F.J.M. acknowledges financial support from Vicerrectorado de Investigación y Desarrollo de la UPV (PAID-05-2009 through project UPV2010-0096). A.M and P. R-H acknowledge the supercomputer time provided by the Red Española de Supercomputación (RES).

**Table I:** Structural parameters of the triclinic and monoclinic structures of $CuWO_4$ and different pressure.

| Structure | Pressure (GPa) | a (Å) | b (Å) | c (Å) | α (º) | β (º) | γ (º) | Volume ($Å^3$) |
|---|---|---|---|---|---|---|---|---|
| $P\bar{1}$ | 0.7 | 4.695 | 5.827 | 4.876 | 91.627 | 92.385 | 83.004 | 132.23 |
| | 16 | 4.523 | 5.716 | 4.783 | 90.831 | 89.910 | 85.566 | 123.26 |
| *P2/c* | 16 | 4.524 | 5.529 | 4.896 | 90 | 90.861 | 90 | 122.44 |

**Table II:** Raman modes, pressure coefficients, and Grüneisen parameters for phase I

| $P\bar{1}$ | | | | | | |
|---|---|---|---|---|---|---|
| | Literature | *Ab initio* | | Present experiment | | |
| Mode (sym) | ω ($cm^{-1}$) 1 bar | ω ($cm^{-1}$) 1 bar | dω/dP ($cm^{-1}$/GPa) | ω ($cm^{-1}$) 1bar | dω/dP ($cm^{-1}$/GPa) | γ |
| $A_g$ | | 88.1 | 1.34 | 95.3 | 1.15 | 2.06 |
| $A_g$ | | 115.6 | 1.84 | 127.6 | 1.65 | 2.21 |
| $A_g$ | | 137.7 | 0.93 | 149.1 | 1.97 | 2.26 |
| $A_g$ | 180 | 164.4 | 1.75 | 179.2 | 1.64 | 1.56 |
| $A_g$ | 192 | 178.0 | 1.19 | 191.0 | 0.94 | 0.84 |
| $A_g$ | 224 | 209.2 | 1.99 | 223.8 | 1.88 | 1.44 |
| $A_g$ | 283 | 263.5 | 1.35 | 282.6 | 1.35 | 0.82 |
| $A_g$ | 293 | 276.2 | 2.03 | 292.6 | 2.28 | 1.33 |
| $A_g$ | 315 | 294.3 | 2.38 | 316.2 | 2.23 | 1.21 |
| $A_g$ | 358 | 341.0 | 3.71 | 358.2 | 1.98 | 0.95 |
| $A_g$ | 398 | 374.9 | 1.45 | 397.5 | 1.72 | 0.74 |
| $A_g$ | 405 | 391.9 | 1.79 | 403.4 | 2.68 | 1.14 |
| $A_g$ | 479 | 454.1 | 3.44 | 479.9 | 5.53 | 1.97 |
| $A_g$ | 550 | 525.2 | 2.38 | 549.8 | 3.19 | 0.99 |
| $A_g$ | 676 | 633.6 | 3.31 | 676.7 | 4.78 | 1.21 |
| $A_g$ | 733 | 695.8 | 2.78 | 733.1 | 2.27 | 0.53 |
| $A_g$ | 779 | 763.2 | 2.42 | 778.9 | 3.93 | 0.86 |
| $A_g$ | 906 | 854.4 | 1.58 | 905.9 | 3.54 | 0.67 |

**Table III.** Raman modes and pressure coefficients for phases II and III.

| | *P2/c* | | | | Phase III | |
|---|---|---|---|---|---|---|
| | *Ab initio* | | Present experiment | | Present experiment | |
| Mode (sym) | ω ($cm^{-1}$) 12.3 GPa | dω/dP ($cm^{-1}$/GPa) | ω ($cm^{-1}$) 12.7 GPa | dω/dP ($cm^{-1}$/GPa) | ω ($cm^{-1}$) 17.1 GPa | dω/dP ($cm^{-1}$/GPa) |
| $B_g$ | 96.6 | 0.90 | 90.3 | 0.75 | 99.7 | 0.18 |
| $A_g$ | 128.9 | -0.09 | 110.8 | 0.32 | 117.5 | 0.45 |
| $B_g$ | 156.9 | 0.87 | 154.3 | 0.19 | 140.6 | 0.36 |
| $B_g$ | 178.2 | 0.45 | 172.7 | 1.01 | 171.2 | -0.16 |
| $B_g$ | 190.2 | 0.43 | 185.6 | 0.39 | 206.6 | 0.13 |
| $A_g$ | 191.9 | 2.50 | 203.1 | 0.38 | 254.8 | 1.23 |
| $A_g$ | 274.9 | 1.34 | 214.5 | 0.44 | 313.7 | 0.71 |
| $B_g$ | 285.2 | 2.45 | 265.1 | 0.67 | 375.0 | 3.91 |
| $A_g$ | 312.2 | 1.48 | 284.8 | 0.09 | 426.9 | 2.93 |
| $B_g$ | 315.9 | 1.58 | 315.3 | 0.29 | 445.2 | 1.35 |
| $B_g$ | 367.3 | 3.57 | 331.4 | 0.31 | 564.3 | 3.21 |
| $A_g$ | 390.5 | 1.33 | 459.3 | 1.43 | 608.8 | 1.70 |
| $B_g$ | 505.4 | 3.30 | 502.4 | 3.57 | 719.4 | 3.83 |
| $A_g$ | 547.8 | 2.97 | 560.6 | 2.57 | 754.2 | 3.47 |
| $B_g$ | 645.2 | 3.74 | 699.4 | 2.78 | 850.1 | 4.74 |
| $A_g$ | 686.0 | 2.99 | 745.2 | 1.77 | 966.2 | 2.33 |
| $B_g$ | 749.1 | 4.04 | 918.7 | 2.12 | | |
| $A_g$ | 847.3 | 3.09 | 963.4 | 2.53 | | |

**Table IV:** Calculated lattice parameters and internal coordinates for $CuWO_4$.

| | $P\bar{1}$ (0.3 GPa) | *P2/c* (10.3 GPa) |
|---|---|---|
| *a* | 4.8352 | 4.5527 |
| *b* | 6.0538 | 5.6479 |
| *c* | 4.9429 | 4.9002 |
| α | 93.26 | 90.00 |
| β | 94.25 | 90.06 |
| γ | 80.11 | 90.00 |
| Cu | 2i (0.7523, 0.5662, 0.6569) | 2f (0.5, 0.6641, 0.25) |
| W | 2i (0.7899, 0.5705, 0.1228) | 2e (0, 0.1802, 0.25) |
| $O_1$ | 2i (0.2663, 0.9023, 0.6204) | 4g (0.2598, 0.3817, 0.4080) |
| $O_2$ | 2i (0.2150, 0.9538, 0.0813) | 4g (0.2251, 0.8940, 0.4302) |
| $O_3$ | 2i (0.5095, 0.7492, 0.3471) | |
| $O_4$ | 2i (0.9777, 0.7434, 0.8318) | |

**Table V. Appendix**

| Mode (sym) | $P\bar{1}$ ω ($cm^{-1}$) 1 atm | dω/dP ($cm^{-1}$/GPa) | Mode (sym) | $P2/c$ ω ($cm^{-1}$) 10.3 GPa | dω/dP ($cm^{-1}$/GPa) |
|---|---|---|---|---|---|
| $A_u$ | 0 | | $B_u$ | 0 | |
| $A_u$ | 0 | | $B_u$ | 0 | |
| $A_u$ | 0 | | $A_u$ | 0 | |
| $A_u$ | 101.4 | 5.09 | $B_u$ | 138.3 | 1.40 |
| $A_u$ | 157.3 | 0.72 | $B_u$ | 200.8 | 0.87 |
| $A_u$ | 214.2 | 0.23 | $A_u$ | 204.8 | 2.70 |
| $A_u$ | 239.0 | -1.64 | $B_u$ | 228.5 | -1.73 |
| $A_u$ | 266.1 | 1.37 | $B_u$ | 258.7 | 1.72 |
| $A_u$ | 281.0 | 0.71 | $B_u$ | 299.9 | -0.23 |
| $A_u$ | 320.5 | 1.27 | $A_u$ | 300.6 | 3.33 |
| $A_u$ | 332.3 | 1.98 | $A_u$ | 332.1 | 1.13 |
| $A_u$ | 383.8 | 4.21 | $A_u$ | 433.3 | 3.01 |
| $A_u$ | 438.9 | 2.67 | $B_u$ | 466.2 | 4.98 |
| $A_u$ | 474.8 | 4.03 | $A_u$ | 478.7 | 5.04 |
| $A_u$ | 516.5 | 1.73 | $B_u$ | 508.6 | 2.67 |
| $A_u$ | 639.6 | 1.51 | $A_u$ | 612.3 | 3.19 |
| $A_u$ | 727.7 | 0.99 | $B_u$ | 692.4 | 3.35 |
| $A_u$ | 852.7 | 0.56 | $A_u$ | 832.9 | 2.46 |

**Figure captions:**

**Figure 1**: Selected X-ray diffraction patterns of the experiment performed using silicone-oil as pressure-transmitting medium. The ticks indicate the identified Bragg reflections of different structures. Distinctive peaks of the wolframite phase are labelled and the emerging peaks related to phase III are indicated with an arrow (at 20.3 GPa).

**Figure 2**: Selected X-ray diffraction patterns of the experiment performed using argon as pressure-transmitting medium. The asterisks indicate the Ar Bragg peak. The (110) and (010) Bragg reflections of the HP wolframite phase are also indicated.

**Figure 3:** LeBail fit of diffraction patterns of phases I (0.7 GPa) and II (16 GPa). Dots: experiments. Lines: model. Ticks indicate the position of calculated Bragg reflections. The lattice parameters and angles at these pressures are given in Table I.

**Figure 4:** Evolution of the normalized lattice parameters with pressure for both experiments: SO (solid symbols) and Ar (empty symbols). The unit-cell parameters are shown as squares (*a*), circles (*b*), and triangles (*c*). The inset shows the pressure dependence of the triclinic angles.

**Figure 5:** Dependence of the normalized unit-cell volume with pressure for both experiments: SO (solid symbols) and Ar (empty symbols). Solid and dashed lines correspond to the EOS obtained from the SO and Ar data, respectively. The inset shows the variation of the normalized FWHM of the (100) Bragg peak of phase I with pressure.

**Figure 6:** Images of a $CuWO_4$ single crystal pressurized at 8.5 and 10.3 GPa in a DAC using silicone oil as pressure transmitting medium. At 8.5 GPa the uniform brownish colour indicates that only one phase is present. At 10.3 GPa the colour domains are due to the subtle phase transition and reveals two-phase coexistence.

**Figure 7:** Raman spectra of $CuWO_4$ recorded at selected pressures. The ticks indicate the experimental modes assignation. At 12.7 GPa the upper ones are those assigned to the triclinic structure while the lower ones are those identified as wolframite P2/c modes. Spectra collected with the LabRam setup.

**Figure 8:** Pressure dependence of the Raman mode frequencies for different phases of $CuWO_4$.

**Figure 9**: *Ab initio* calculated energy vs. volume curves for the triclinic and wolframite-type structures. The inset shows the enthalpy difference between both structures.

**Figure 1**

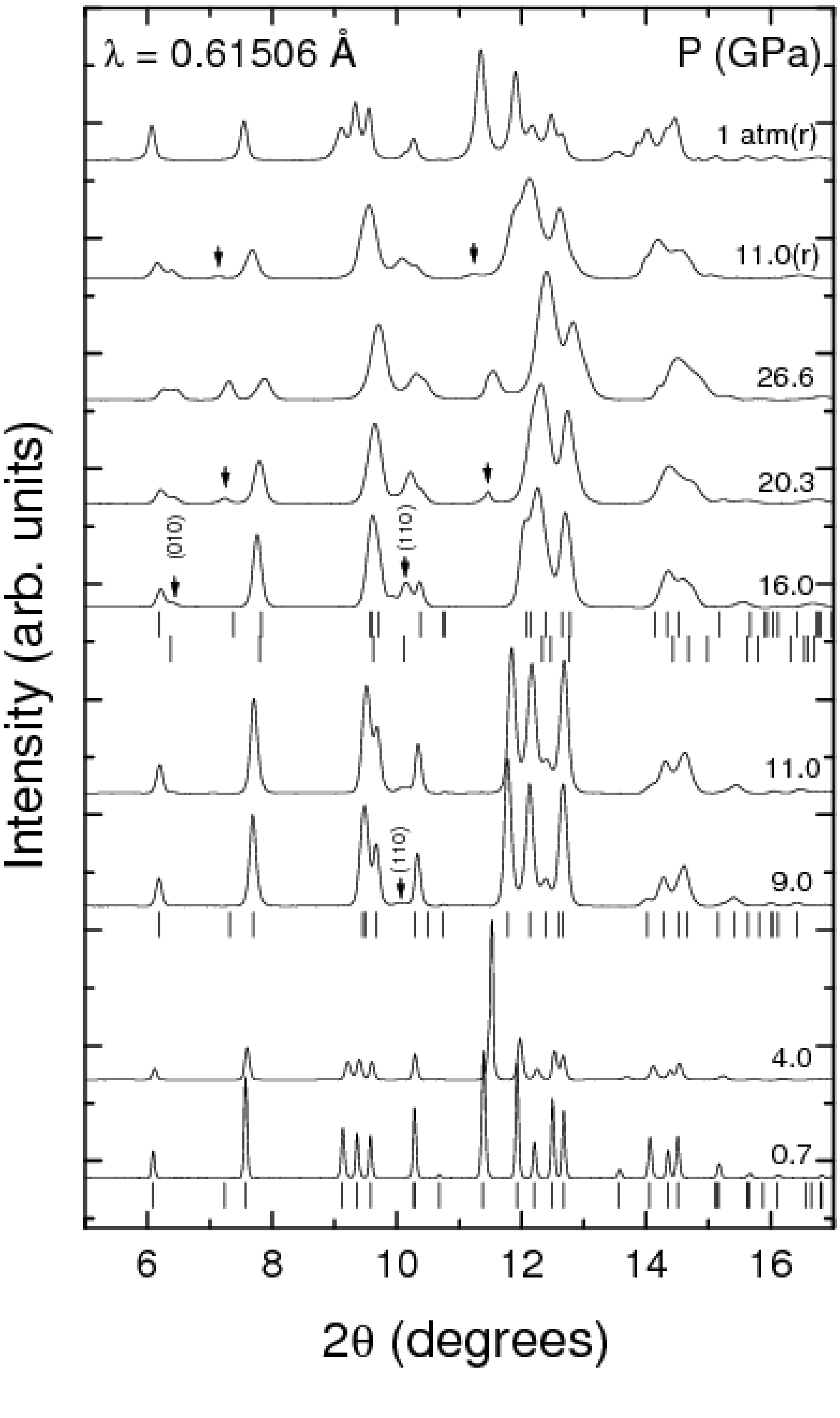

**Figure 2**

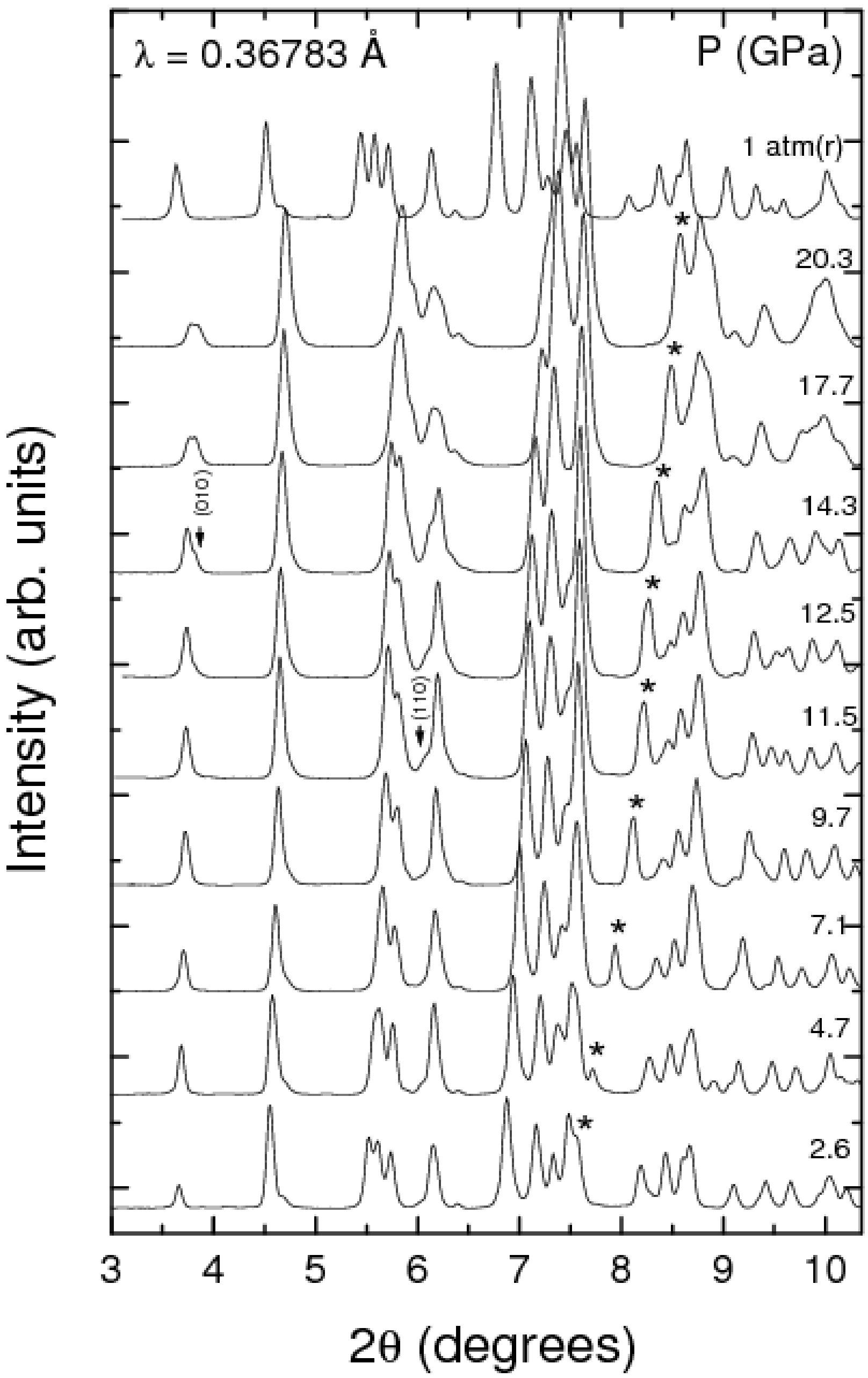

**Figure 3**

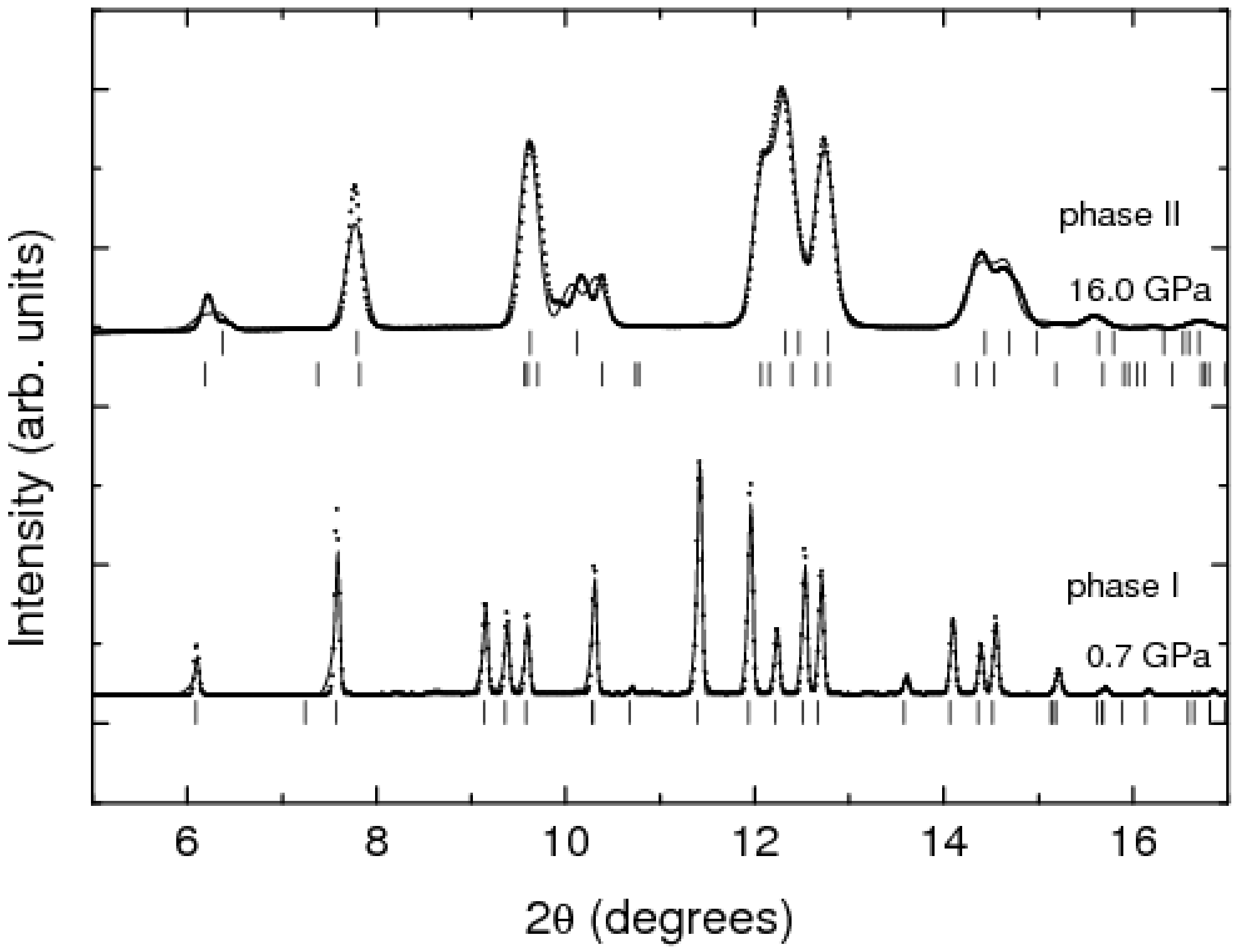

**Figure 4**

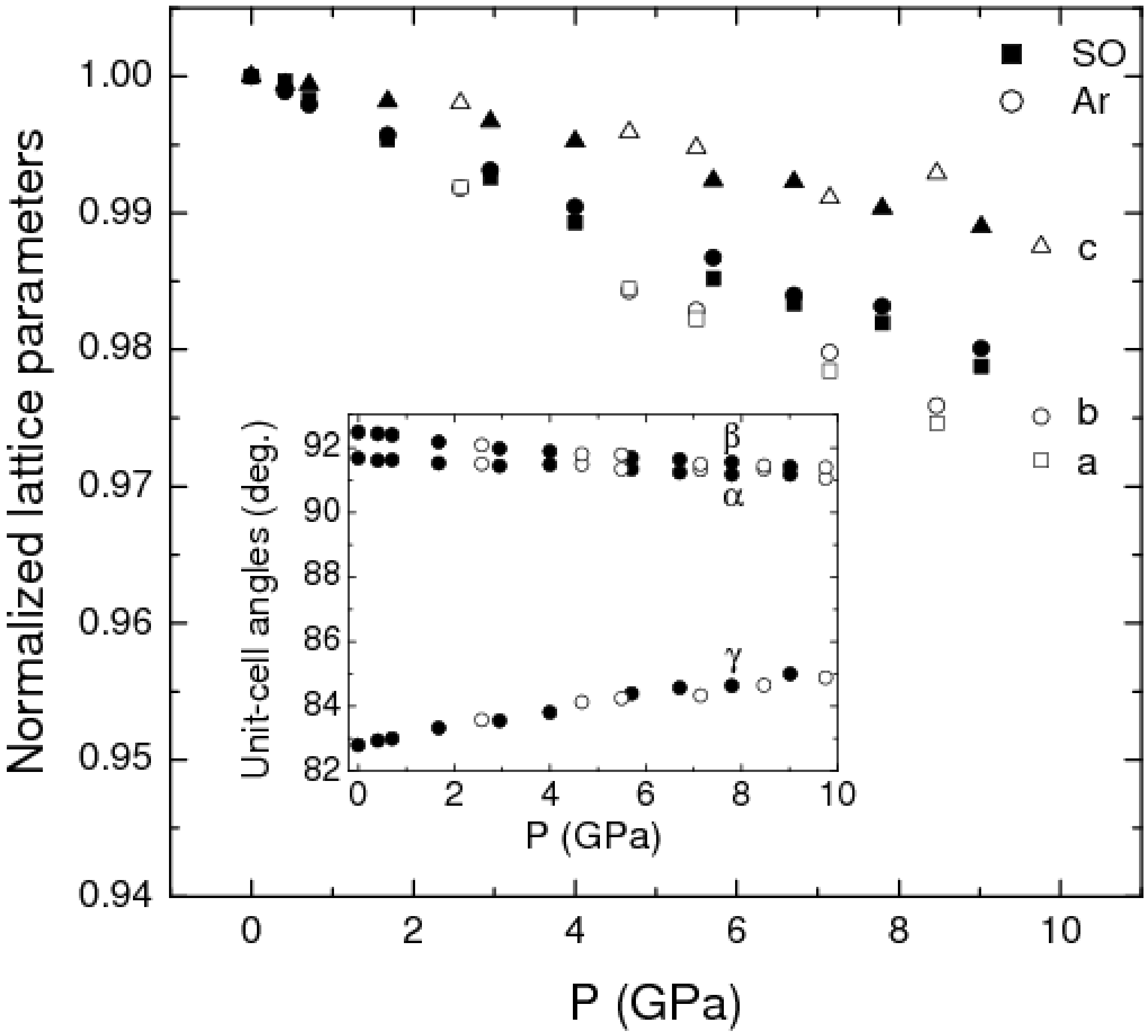

**Figure 5**

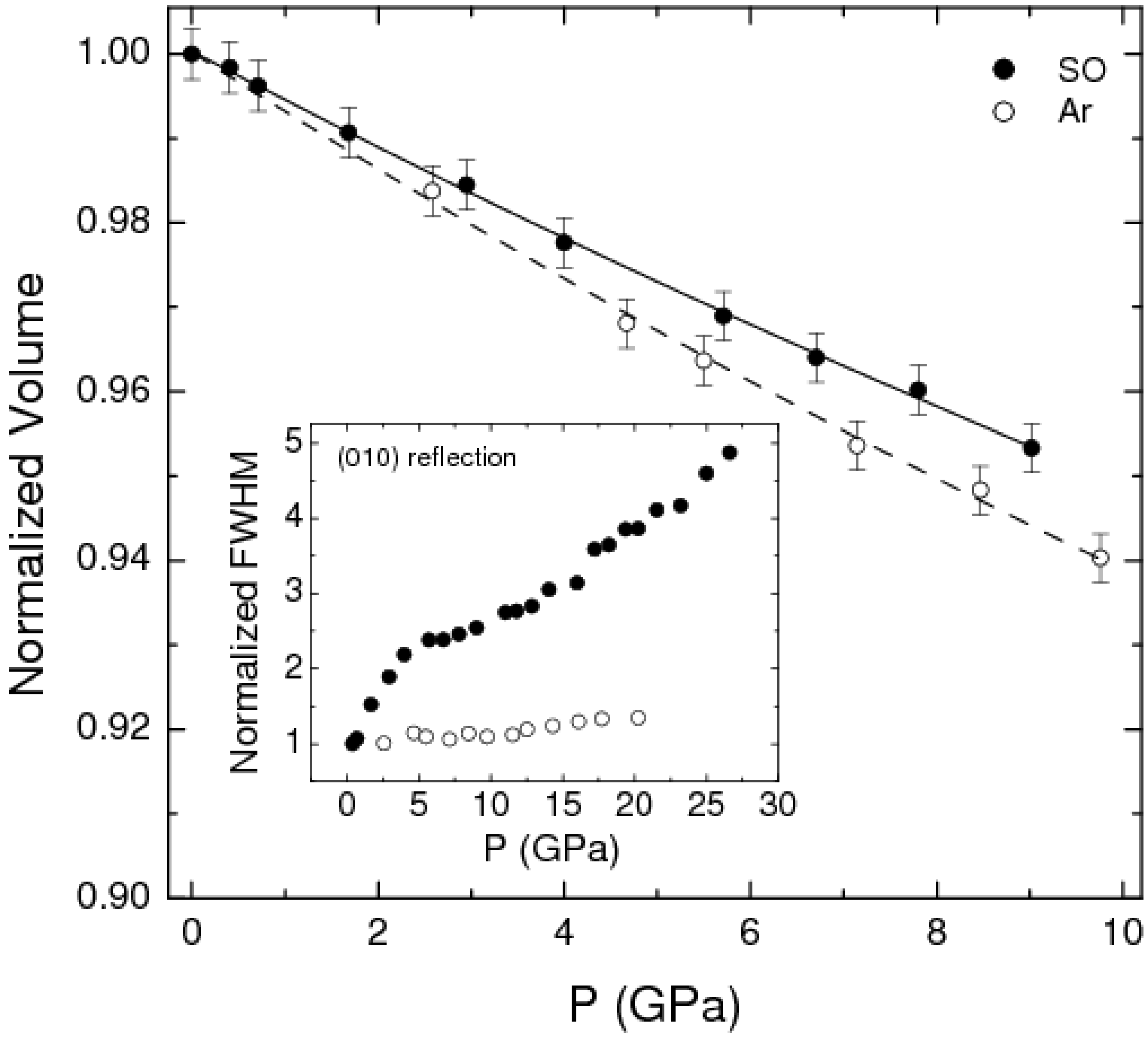

**Figure 6**

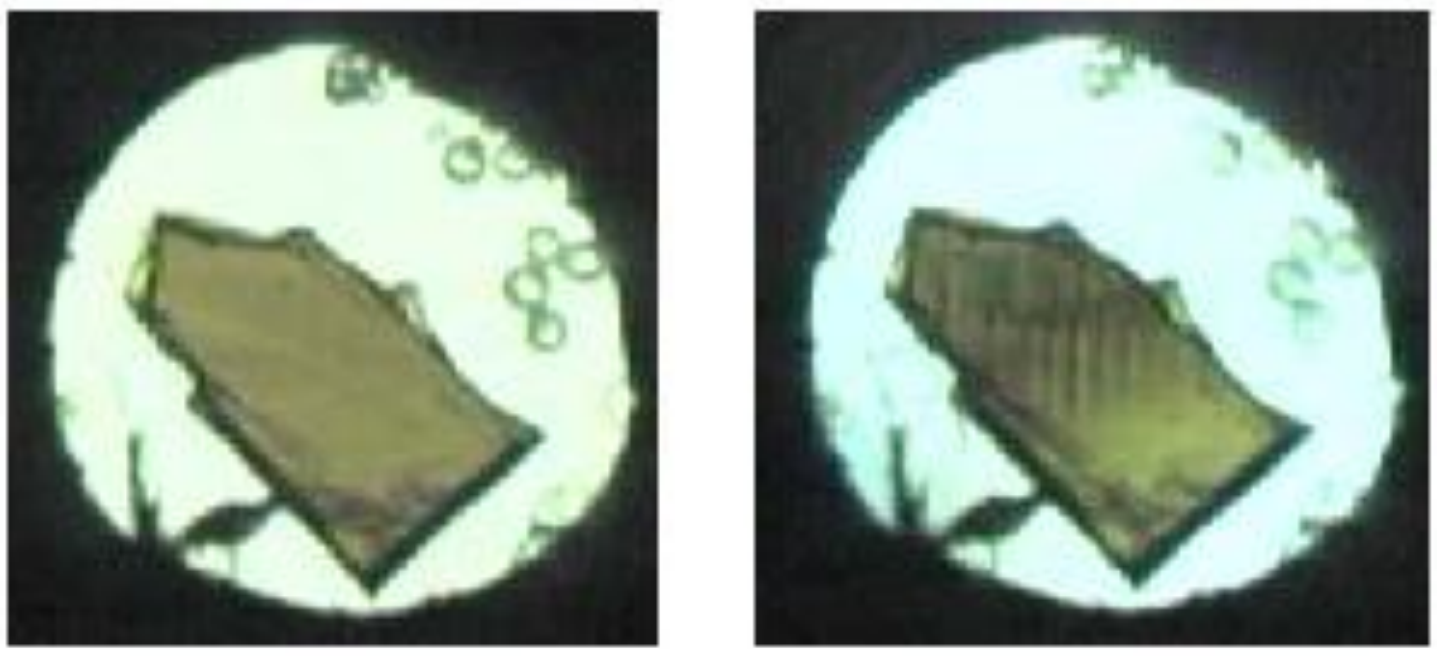

**Figure 7**

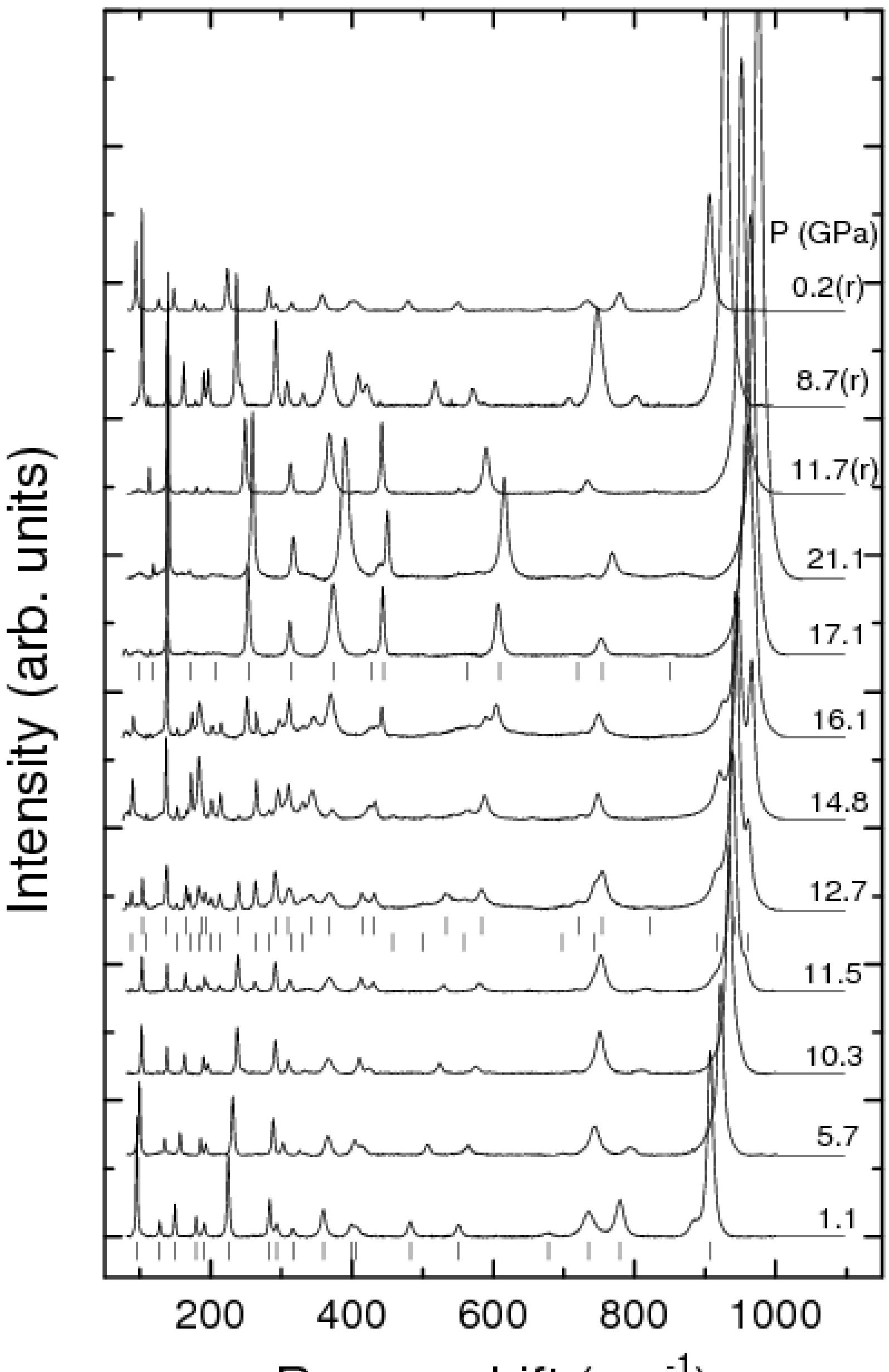

**Figure 8**

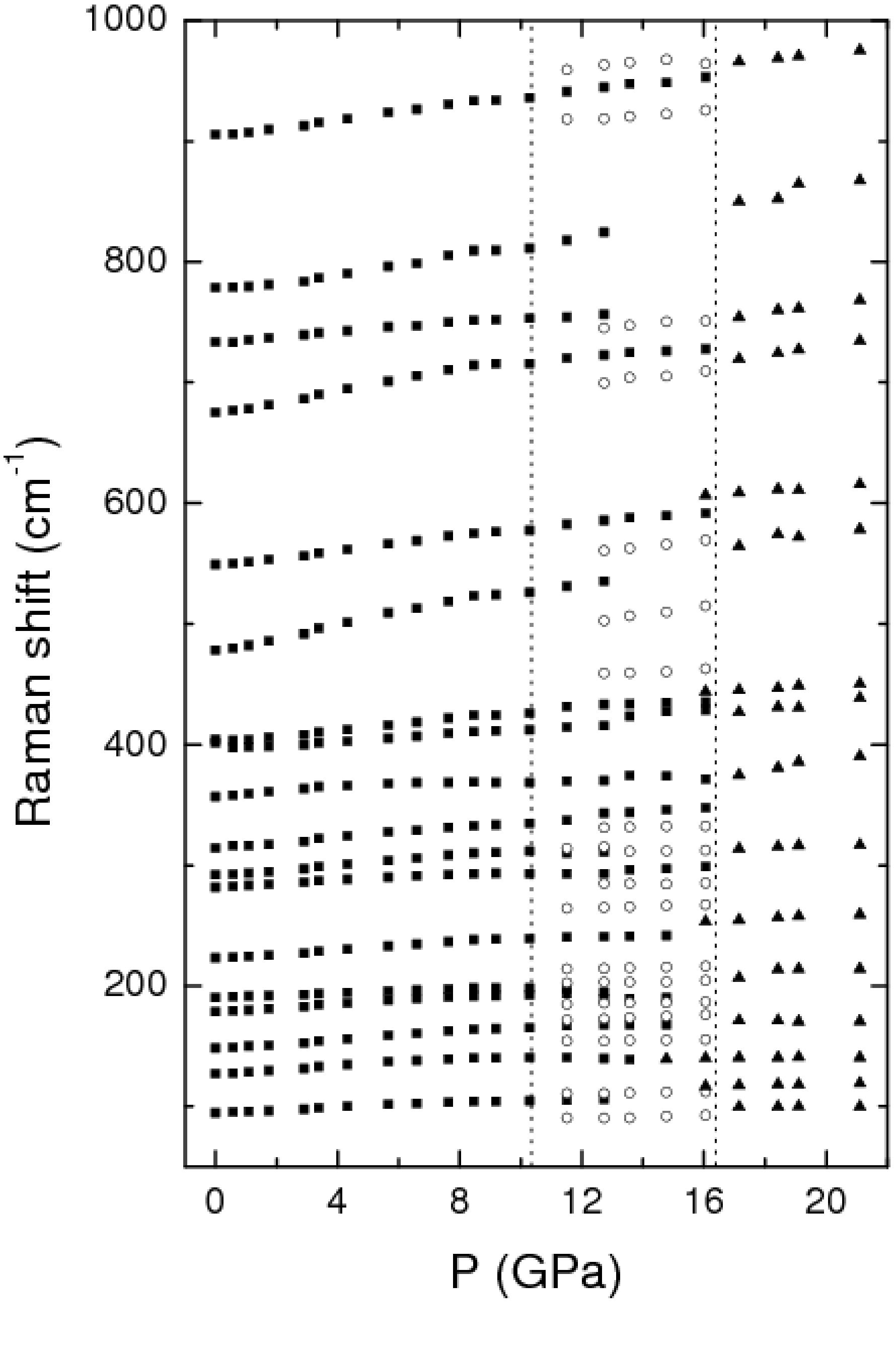

**Figure 9**

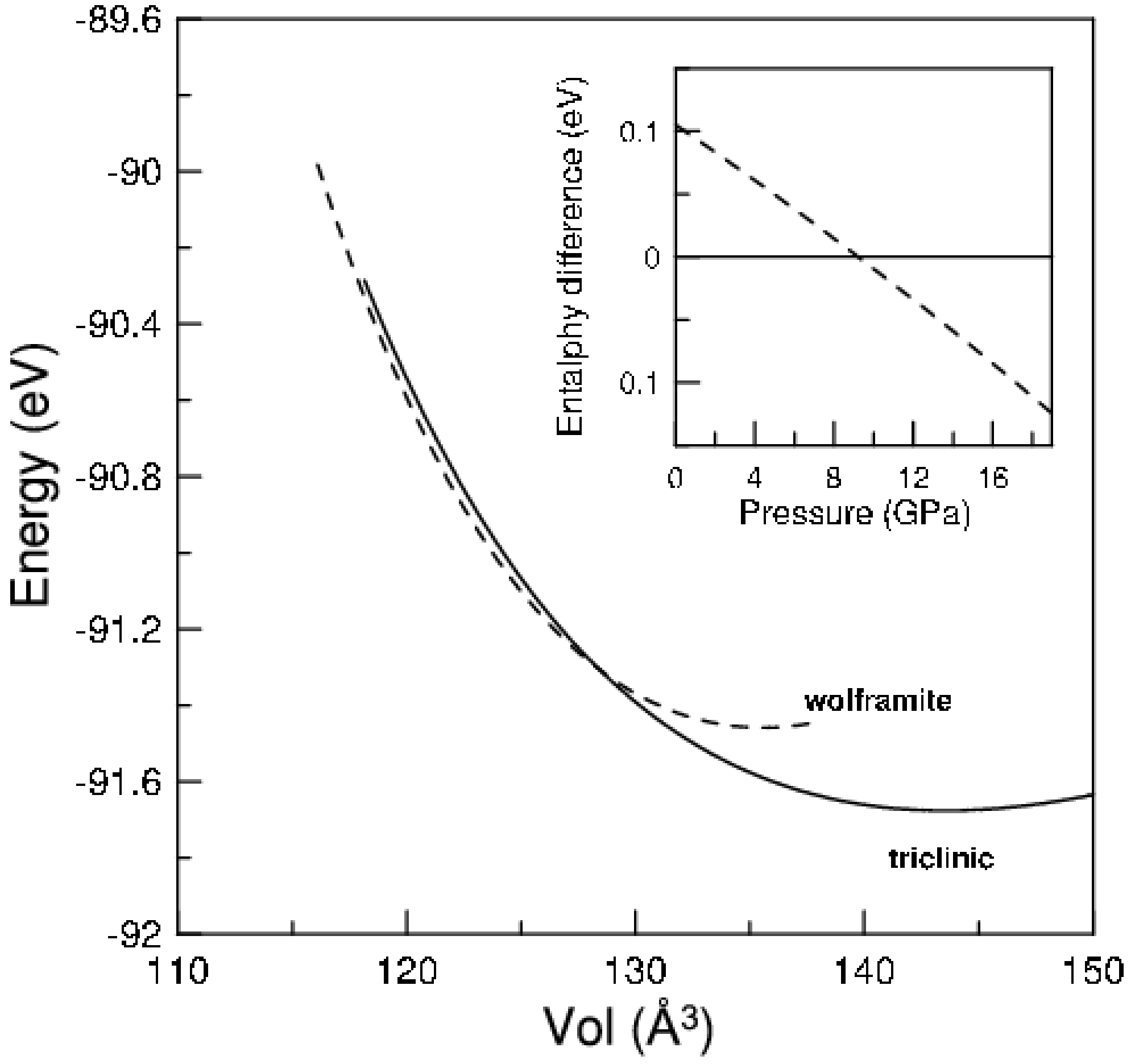